\documentclass[twocolumn,showpacs,10pt,amsmath,amssymb]{revtex4}
\usepackage{graphicx}

\usepackage{amsmath,amsfonts,amssymb,bm}

\begin{document}

\draft
\title{Escape process in systems characterised by stable noises and position-dependent resting times}

\author
{Tomasz Srokowski}

\affiliation{
 Institute of Nuclear Physics, Polish Academy of Sciences, PL -- 31-342
Krak\'ow,
Poland }

\date{\today}

\begin{abstract}
Stochastic systems characterised by a random driving in a form of the general stable noise are considered. The particle experiences 
long rests due to the traps the density of which is position-dependent and obeys a power-law form attributed to the underlying 
self-similar structure. Both one and two dimensional case are analysed. 
The random walk description involves a position-dependent waiting time distribution. On the other hand, 
the stochastic dynamics is formulated in terms of the subordination technique where the random time generator is position-dependent. 
The first passage time problem is addressed by evaluating a first passage time density distribution and an escape rate. 
The influence of the medium nonhomogeneity on those quantities is demonstrated; moreover, the dependence of the escape rate on the stability 
index and the memory parameter is evaluated. Results indicate essential differences between the Gaussian case and the case involving 
L\'evy flights. 
\end{abstract} 


\maketitle


\section{Introduction}

The continuous time random walk theory (CTRW) is a natural approach to the dynamics of the complex systems which are characterised 
by a large number of strong interacting components, a sensitivity on the initial conditions, resulting in an apparent stochastic 
motion, as well as long memory effects. CTRW resolves itself to a single jump probability density function (PDF) $\psi(l,\tau)$ 
which determine both a length $l$ of the jump and a time interval $\tau$ between successive jumps. In the simplest case, CTRW is decoupled: 
$\psi(l,\tau)=\lambda(l)w(\tau)$. If $w(\tau)$ possesses algebraic tails, $\tau^{-1-\beta}$ ($0<\beta<1)$, 
the memory effects emerge: PDF obeys a fractional Fokker-Planck equation and the transport is subdiffusive \cite{met}. 
In the L\'evy walk model \cite{zab}, in turn, the jump length is related to the velocity which is finite and makes the jumping time 
finite, then $\psi(l,\tau)=\lambda(l|\tau)w(\tau)$. In the present paper, we consider another 
version of CTRW assuming that the waiting time PDF depends on the current process value $x$, $w=w(\tau|x)$, 
while jumps are instantaneous \cite{kam}. This ansatz traces back to a picture of a motion in a medium with nonhomogeneously 
distributed traps which may invoke periods of long rests; such a picture is common for complex \cite{tak} 
and disordered systems \cite{bou}. Moreover, in some dynamical systems (in particular of a generalised Lorentz gas type) 
the trajectories can stick to stability islands, nonhomogeneously distributed in the phase space, 
and abide there for a long time (dynamical traps) \cite{zas}. The presence of a position-dependent distribution of sojourns
is obvious for problems related to a population movement \cite{bro1} since some area provide more opportunities than the other 
whether in respect to an employment offer or as a touristic attraction. 
Despite ubiquity of problems involving the position-dependent memory, the CTRW taking into account that effect has been rarely 
investigated. The conditional form of $w(\tau|x)$ leads to a variable, position-dependent, diffusion coefficient 
in a corresponding Fokker-Planck equation \cite{sro06} and is a source of the anomalous transport: both a subdiffusion and 
an enhanced diffusion. The anomalous transport emerges even if $w(\tau|x)$ is Poissonian, 
i.e. in the absence of any memory effects. 

The analysis of many physical, in particular complex, systems may require taking into account long jumps by generalising the Gaussian form 
of $\lambda(l)$ to arbitrary L\'evy stable distributions (L\'evy flights). The $\alpha$-stable form of $\lambda(l)$ may 
reflect a self-similar medium structure in complex systems where clusters of short jumps coexist with long jumps at all scales. 
The L\'evy flights occur in many areas of science \cite{shles,barn} and, in particular, are observed in 
heterogeneous materials \cite{tall}. The space heterogeneity may require, in turn, 
the introduction of the variable diffusion coefficient. This happens, for example, in the folded polymers \cite{bro}, and 
in a random walk description of transport in a composite medium with many layers characterised by a position-dependent trapping 
time statistics \cite{sti}. 

The decoupled CTRW can alternatively be formulated in the framework of a subordination technique \cite{subor} and 
generalised to the case of a medium with heterogeneously distributed traps where the heterogeneity is taken into account as a variable 
trap density \cite{sro14,sro15}. Then the problem resolves itself to a fractional Fokker-Planck equation with a position-dependent 
diffusion coefficient. 

When a system is restricted in space, the particle, after reaching a boundary, may either escape or return to 
the bulk; then appropriate boundary conditions apply: the absorbing and/or reflecting ones. The escape problem is well-known 
in the Gaussian case \cite{han}; it was also discussed for the L\'evy flights when trajectories are discontinuous and 
the boundary conditions nonlocal which results in a nontrivial leapover statistics \cite{kore}. 
The escape problem for the L\'evy flights was studied both for a free particle \cite{dyb} and in the presence of an external 
deterministic force \cite{dyb1,szcz}. It was demonstrated in the field of the disordered systems that the first passage time 
statistics may be different for the annealed and quenched disorder \cite{luo}. 
The first passage time distribution has divergent moments, in particular a mean, for waiting time distributions with the 
algebraic tails \cite{yus}. 

The time characteristics of the escape process must be influenced by the medium structure since 
the particle abides longer in the region where the trap density is large. Indeed, the variance for 
the confined L\'evy flights rises slowly with time if this density rises fast with the distance \cite{sro15}. In the present paper, 
we study the influence of the medium heterogeneity on the escape process on the assumption that the jump length is governed 
by the symmetric stable L\'evy distribution and the sojourn time by the one-sided stable distribution, i.e. they have the algebraic tails. 
The paper is organised as follows. In Sec.II we define CTRW process and the corresponding Langevin dynamics. 
Sec.III is devoted to the first passage time problem: In Sec.IIIA we demonstrate how the escape rate depends on the trap density, 
as well as on the parameters $\alpha$ and $\beta$. The first passage time PDF is evaluated in the framework of the subordination 
technique in Sec.IIIB.

\section{Random walk with a position-dependent waiting time and the Langevin equation} 

We define the random walk process by two density distributions: the jump length and the waiting time distribution. 
They describe movement of a particle performing jumps and resting between successive jumps while the jumps 
themselves are instantaneous \cite{uwa}. 
The first distribution determines a probability $Q({\bf l})d^d{\bf l}$ ($d$ is a dimension of the system) 
that the displacement assumes a value in the interval $({\bf l},{\bf l}+d{\bf l})$. 
We assume that $Q({\bf l})$ is a general symmetric stable distribution and the components of ${\bf l}$ are independent. Then the 
characteristic function reads 
\begin{equation}
\label{q}
\widetilde Q({\bf k})=\exp(-\sum_{i=1}^d|k_i|^\alpha),
\end{equation}
where $0<\alpha\le2$. 
In the case of the L\'evy flights, the divergence of the mean squared displacement poses a difficulty if we are dealing with dynamics 
of massive particles. However, one may take into account that the finiteness of the system imposes a natural cut-off and makes the variance 
actually finite. This can be formalised either by introducing truncated distributions \cite{man} or a multiplicative noise at the boundary \cite{sro15}. 
On the other hand, the infinite variance does not violate physical principles for such physical problems as diffusion in the energy space 
in spectroscopy or diffusion on a polymer chain in chemical space \cite{met}. 

The waiting time distribution $w(\xi)$, in turn, depends on the current position and this dependence reflects a nonhomogeneous trap distribution. 
Two cases are distinguished. If $w(\xi)$ has a Poissonian form, the process is Markovian without any memory effects \cite{kam} while 
the presence of those effects implies algebraic and long tails of $w(\xi)$. They emerge, for example, 
in disordered systems as a result of the exponential distribution of site energies \cite{luo}. 
We assume $w(\xi)$ as a one-sided completely asymmetric L\'evy distribution $L_\beta(\xi)$ 
with a skew parameter $\beta$, $0<\beta<1$; it has the asymptotics $\propto \xi^{-1-\beta}$ and for small $\xi$ obeys a form \cite{schn}, 
\begin{equation}
\label{lev1s}
w(\xi)\propto\xi^{-c_1}\hbox{e}^{-c_2\xi^{-c_3}},
\end{equation}
where $c_1=(2-\beta)/2(1-\beta)$, $c_2=(1-\beta)\beta^{\beta/(1-\beta)}$ and $c_3=\beta/(1-\beta)$. The stable form of 
the distribution $w(\xi)$ is consistent with a subordination formalism which we will introduce at the end of this section. 
The position dependence enters $w(\tau|{\bf r})$ via a trap density $g({\bf r})$. It is natural 
to expect that a large trap density results in a long waiting time: 
the larger $g({\bf r})$ the longer the time the particle abides in the vicinity of ${\bf r}$. Therefore we put 
\begin{equation}
\label{wodt}
w(\tau|{\bf r})=L_\beta(\tau/g({\bf r}))  
\end{equation}
and the equations defining the CTRW process read, 
\begin{eqnarray}
\label{eqrw}
{\bf r}_{n+1}&=&{\bf r}_n+\boldsymbol\xi_Q\nonumber\\
t_{n+1}&=&t_n+g({\bf r}_n)\xi_w, 
\end{eqnarray}  
where the components of $\boldsymbol\xi_Q$ and $\xi_w$ stand for mutually independent and uncorrelated random numbers distributed 
according to $Q(\boldsymbol\xi)$ and $w(\xi)$, respectively. If $w(\xi)$ has a Poissonian form, Eq.(\ref{eqrw}) is equivalent to 
a master equation the solution of which, $p_0(x,t)$, satisfies, in a limit of small wave numbers, a fractional Fokker-Planck equation 
with a variable diffusion coefficient \cite{sro06,sro15}. The density for $\beta<1$, $p(x,t)$, follows from a simple scaling 
of the time variable which produces a Laplace transform ${\widetilde p}(x,s)=s^{\beta-1}{\widetilde p}_0(x,s^\beta)$ \cite{met}; then the time derivative 
in the Fokker-Planck equation becomes fractional (cf. Eq.(\ref{glef})). 
We assume a power-law form of the trap density $g({\bf r})$ which in the one-dimensional case reads 
\begin{equation}
\label{godx}
g(x)=|x|^\theta, 
\end{equation}
where $\theta$ is a parameter characterising a degree of the medium nonhomogeneity and may assume, in principle, both positive and negative 
values. $g(x)$ in the form (\ref{godx}) may be interpreted as a result of a fractal trap structure. Then the fractal density 
is proportional to $|x|^{d_f-1}$, where $d_f$ stands for 
a fractal (capacity) dimension \cite{lich,hav}. More precisely, one evaluates, for a fixed, high resolution, the number of traps 
up to the distance $x$ ($|x|<1$) and normalises the result to the number of traps in the unit interval. 
The comparison of the above expression for the fractal density with Eq.(\ref{godx}) relates the parameter $\theta$ 
to the fractal dimension, $\theta=d_f-1$, and imposes a restriction on $\theta$: $-1\le\theta\le0$. Therefore, the 
interpretation of Eq.(\ref{godx}) as the fractal density implies that $\theta$ must always be non-positive. 
In the following, we will also consider two versions of a two-dimensional case, ${\bf r}=(x,y)$. First,  
we assume an isotropic trap distribution for which  
\begin{equation}
\label{godr}
g({\bf r})=r^\theta, 
\end{equation}
where $r=|{\bf r}|$, and the fractal density equals $r^{d_f-2}$; accordingly, $\theta=d_f-2$ and $-2\le\theta\le0$. 
As a second case, we consider a nonisotropic trap distribution, namely a structure where the nonhomogeneity 
is restricted to the one direction, i.e., $g({\bf r})$ obeys Eq.(\ref{godx}) in the $x$- direction while $g(y)=1$. 
We will call this case 'nonisotropic'. Dimension of the corresponding fractal is just a sum of dimensions in both 
directions: $d_f=d_f^x+d_f^y=\theta+2$ ($-1\le\theta\le0$). Therefore, $d_f$ depends on $\theta$ in the same way 
for the isotropic and nonisotropic cases but the range of $\theta$ is different \cite{met1}. 
\begin{center}
\begin{figure}
\includegraphics[width=95mm]{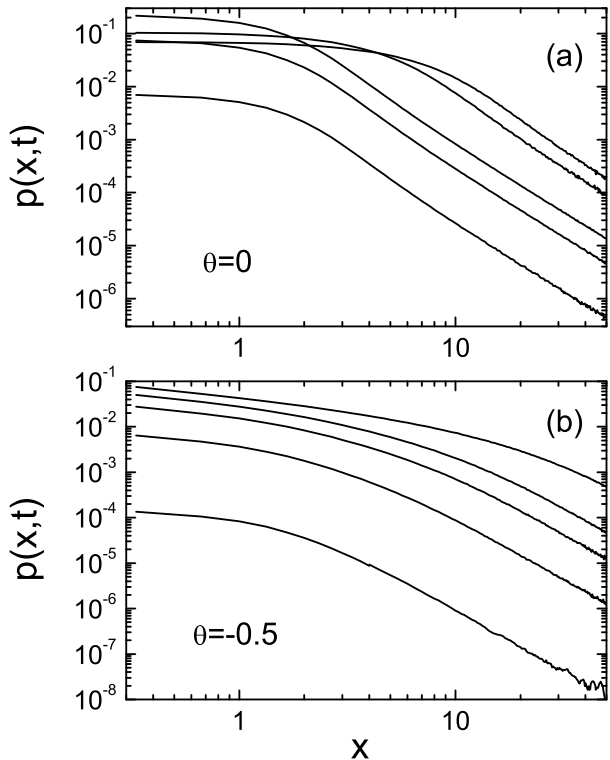}
\caption{Time evolution of the CTRW density calculated for $\alpha=1.5$, $\beta=0.9$, $x_0=0.1$ and two values of $\theta$. 
The curves (from bottom to top) correspond to the following times: 5, 6, 10, 50 and 100 (a); 450, 500, 550, 600 and 800 
(b).}
\end{figure}
\end{center} 

The influence of the variable trap density on the PDF time evolution is illustrated in Fig.1. The one-dimensional trajectories 
were evaluated by sampling the jump length $l$ and the waiting time $\tau$ from the stable distributions $Q(l)$ and 
$L_\beta(\tau/|x|^\theta)$, respectively, according to a well-known procedure \cite{wer}. The particle was initially positioned  
at $x=x_0$ and this point was regarded as a trap, namely the first jump took place after a period of resting. Consequently, 
$p(x,t)$ exhibits a peak at the origin which is larger for $\theta<0$ and then the evolution proceeds much slower. 
The tails of all the distributions obey the form $|x|^{-1-\alpha}$. 

Alternatively, the dynamics may be formulated in terms of the subordination technique which is equivalent to CTRW in the diffusion limit. 
The second equation (\ref{eqrw}) governs the random waiting time while the first equation is responsible for 
displacements as a function of a step number $n$. In the continuous limit, the latter equation turns into a Langevin 
equation and $n$ becomes an operational time. The random, physical time $t$ is defined by a continuous counterpart 
of the second equation (\ref{eqrw}) that becomes an adjoined Langevin equation. In the case of the variable trap density, 
the random time generator depends on the current position which explicitly enters the second equation. 
Then the set of the Langevin equations reads \cite{sro14,sro15} 
\begin{eqnarray}
\label{lan}
d{\bf r}(\tau)&=&\boldsymbol\eta(d\tau)\nonumber\\
dt(\tau)&=&g({\bf r})\xi(d\tau). 
\end{eqnarray}  
In the above equations, the increments of the noise $\xi$ represent the one-sided completely asymmetric L\'evy distribution $L_\beta(\xi)$ 
and $\boldsymbol\eta$ refers to a symmetric L\'evy stable distribution with a stability index $\alpha$ which is the same for all directions. 
The second Eq.(\ref{lan}) takes into account the medium structure: longer trapping times are more probable if the trap density 
$g({\bf r})$ is larger. In particular, for $g({\bf r})$ in the form (\ref{godr}) and negative $\theta$, 
the trapping time becomes infinite in the origin. Eq.(\ref{lan}) is equivalent to the following system of equations \cite{sro14,kaz}, 
\begin{eqnarray}
\label{lanm}
d{\bf r}(\tau)&=&g({\bf r})^{-\beta/\alpha}\boldsymbol\eta(d\tau)\nonumber\\
dt(\tau)&=&\xi(d\tau). 
\end{eqnarray}  
Due to the transformation from (\ref{lan}) to (\ref{lanm}), we can describe the dynamics as the ordinary subordination to the random time 
of a multiplicative process in the It\^o interpretation. Then the subordinator, given by the second equation (\ref{lanm}), 
becomes independent of a specific path and represents a renewal process that connects the times $\tau$ and $t$, 
\begin{equation}
\label{num3}
\tau(t)=\hbox{inf}\{\tau: t(\tau)\ge t\}.   
\end{equation} 
As a consequence, PDF of the process ${\bf r}(t)$ can be evaluated from PDF of ${\bf r}(\tau)$ 
by integration over $\tau$. 
The asymptotics of PDF that follows from Eq.(\ref{lanm}) is either a stretched-Gaussian ($\alpha=2$) \cite{sro15}, 
\begin{equation}
\label{solga}
p(x,t)\sim |x|^{\theta}t^{-\frac{\beta(1+\theta\beta)}{(2+\theta\beta)(2-\beta)}}
\exp\left[-A|x|^{(2+\theta\beta)/(2-\beta)}/t^{\frac{\beta}{2-\beta}}\right],
\end{equation} 
where $A=(2/\beta-1)/[\beta^{3/(2-\beta)}(2+\theta\beta)^{2/(2-\beta)}]$, or a power-law ($\alpha<2$), 
\begin{equation}
\label{asyt}
p(x,t)\propto t^{\beta/(\alpha+\theta\beta)}|x|^{-1-\alpha}. 
\end{equation}
Eq.(\ref{lanm}) corresponds to a fractional Fokker-Planck equation where the fractional derivatives are taken 
over both time and position \cite{sro15a}. 
The first passage time problem resolves itself to a solution of that equation with appropriate boundary conditions. 

\section{First passage time statistics}

We put a particle at ${\bf r}={\bf r}_0$ and ask about a time needed to leave a given area. This escape problem is 
described by a first passage time density distribution defined as a probability that the time the particle needs
to reach the edge of the area for the first time lies within the interval $(t,t+dt)$ \cite{red}. 
The problem is determined by a density distribution $p({\bf r},t)$ which satisfies the initial condition 
and the absorbing boundary conditions at the edge. The survival probability, namely the probability 
that the particle never reached those barriers up to time $t$ (it still remains inside the bulk), 
is given by $S(t)=\int p({\bf r},t)d^d{\bf r}$ where the integration is performed over the area inside the barriers. 
The first passage time density distribution reflects the change of the survival probability with time, 
\begin{equation}
\label{pfp}
p_{FP}(t)=-dS(t)/dt, 
\end{equation}
and the mean first passage time, $\int_0^\infty tp_{FP}(t)dt$, 
serves as an estimator of a speed of the escape process. However, if the asymptotics of the waiting time distribution has power-law tails, 
$w(t)\propto t^{-1-\beta}$ with $0<\beta<1$, the mean first passage time diverges. Instead, one can simply characterise 
the first passage time statistics by a median location or an interquantile width \cite{szcz}. 
Alternatively, one may introduce fractional moments of $p_{FP}(t)$, 
\begin{equation}
\label{mom}
M_\delta=\int_0^\infty t^\delta p_{FP}(t)dt, 
\end{equation}
which are finite for $-1\le\delta<\beta$: the right inequality results from the asymptotics of $p_{FP}(t)$ and the left equality follows from 
$p_{FP}(0)=0$ (the particle rests at the initial point). The moment $M_{-1}$ is distinguished, it allows us to substitute the divergent 
mean first passage time $\langle t\rangle$ by the escape rate which we define as an average over the inverse first passage time. Then the quantity 
\begin{equation}
\label{nu1m}
\nu=M_{-1}
\end{equation}
can serve as a measure estimating how fast the particles leave the bulk: large values of $\nu$ mean the fast escape. 
For a given distance between the initial point and the boundary ($L_b$), $L_b\nu$ is just a mean (effective) particle velocity 
in the bulk. $\nu$ has been recently applied as a measure for the search efficiency in a study 
of random search processes involving long jumps \cite{paly}. For the L\'evy flights, the arrival at the barrier often means 
a substantial overleap but, since jumps are instantaneous, the position of the particle 
after a jump does not influence the escape time and the mean velocity. 
In the following, we estimate the speed of the escape process by evaluating the rate $\nu$. 
The fractional moments corresponding to positive $\delta$ have a similar meaning as the mean first passage time, 
they will be evaluated in Sec.IIIB.

\subsection{Escape rate derived from CTRW}

The time evolution of the stochastic trajectories is given by Eq.(\ref{eqrw}); the successive displacements ${\bf r}_i$ 
and waiting times $\tau_i$ are sampled from the respective distributions. Since jumps are instantaneous, the escape time $t$ 
is determined by a sum of the waiting times, $t=\sum_{i=1}^N\tau_i$, where $N$ satisfies a condition 
$|\sum_{i=1}^{N+1}{\bf r}_i|\ge L$. 
In the nonhomogeneous case, the density of traps diminishes with the distance and one can expect that the escape time is larger 
when one starts closer to the origin; consequently, the choice of the initial condition influences the results. 
In the following, if not stated otherwise, we assume that the initial position of the particle is uniformly distributed 
inside either an interval (1D case) or a circle (2D case) of a radius $x_0$ and the evolution begins from resting 
at the initial point. The absorbing barrier we define either as two points, $x=\pm L$, or a circle 
of the radius $L$ around the origin. Notice, however, that the boundary conditions corresponding to the absorbing barriers 
must be defined differently for the L\'evy flights than those involving only short jumps \cite{zum}. In fact, they are nonlocal: 
the barrier is reached when $|{\bf r}|\ge L$. The speed of escape is estimated by the rate $\nu$ which has been evaluated 
as a function of the system parameters $\theta$, $\alpha$ and $\beta$; the results are presented in Figs. 2-4. 
\begin{center}
\begin{figure}
\includegraphics[width=95mm]{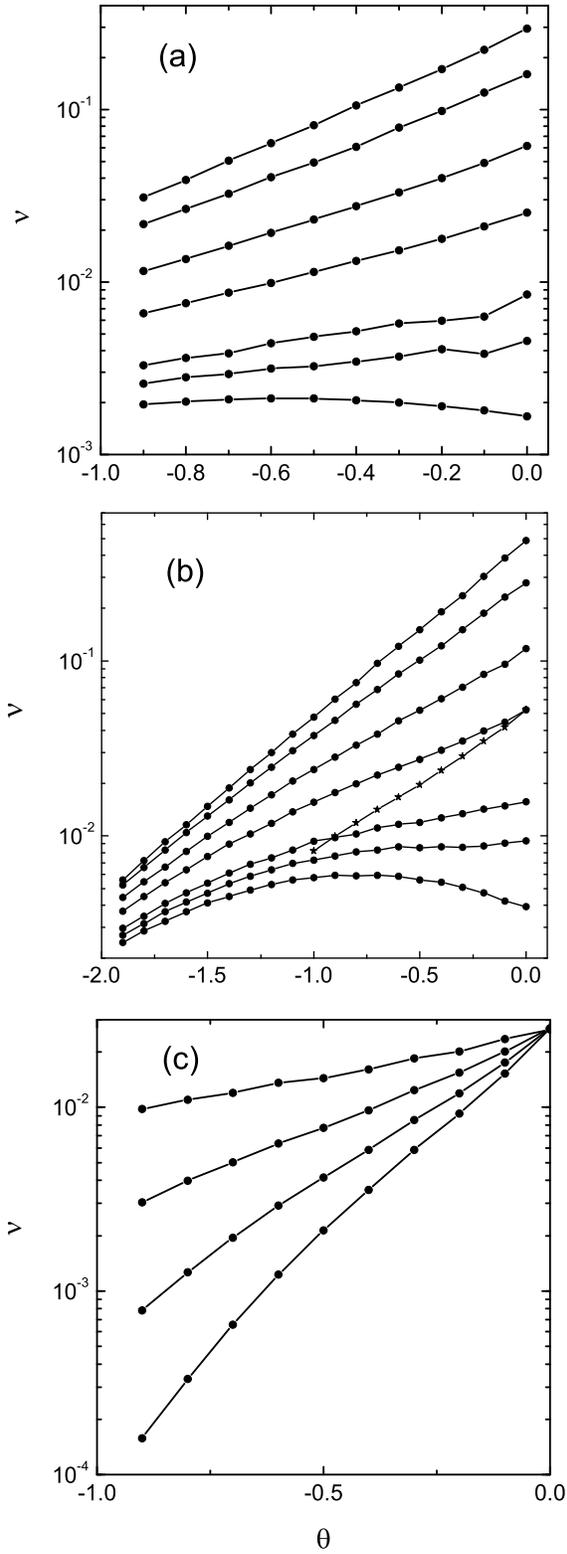}
\caption{(a) and (b): $\nu$ as a function of $\theta$ for $\beta=0.5$, $L=10$ and the following values of 
$\alpha$: 2, 1.9, 1.8, 1.5, 1.2, 0.8 and 0.5 (from bottom to top). Those panels correspond to 1D and 2D cases, 
respectively; stars mark the nonisotropic case. (c): $\nu(\theta)$ calculated for 1D case with the initial 
condition $x_0=10^{-1}$, $10^{-2}$, $10^{-3}$ and $10^{-4}$ (from top to bottom).}
\end{figure}
\end{center} 

Fig.2 presents $\nu$ as a function of $\theta$ for some values of $\alpha$ and the range of $\theta$ in the plot corresponds 
to a fractal with the dimension $0<d_f\le d$, where $d$ is either 1 or 2. According to the upper panel, corresponding to 1D case, 
$\nu$ rises with $\theta$ with an exponential rate for all $\alpha$ except $\alpha=2$ and the slope is larger for smaller $\alpha$. 
That growth can be explained by a strong trapping 
of the particle near the origin for large negative $\theta$. In contrast to that picture, for $\alpha=2$ $\nu(\theta)$ diminishes when 
$\theta$ approaches zero which effect can be attributed to a complicated structure of the trajectory $x(t)$: it consists of many short jumps 
and may return to the origin many times; we shall present such a trajectory in the next section. In 2D case 
we consider two trap geometries: the isotropic distribution and the case when $g({\bf r})$ depends only on $x$, 
according to its definition in Sec.II. Results for the former geometry are presented in the middle panel of Fig.2. The curves exhibit 
a similar pattern as for 1D case but $\nu$ is always smaller for a given $\theta$ which reflects the fact that particle 
easier leaves the neighbourhood of the origin -- which region bear the main responsibility for trapping -- when 
the medium dimensionality is larger. If the trap density varies 
only in one direction (nonisotropic case), the slope of $\nu(\theta)$ is larger, as it is demonstrated in Fig.2 for $\alpha=1.5$. 
Then the trapping is efficient since the particle may remain in the region of a large $g(x)$ (small $|x|$) during its 
movement along trajectories parallel to the $y$ axis. As already mentioned, the results for the nonhomogeneous case may depend on 
the initial conditions. This observation is illustrated in the lower part of Fig.2 where $\nu$ was calculated for various 
but fixed initial conditions, instead of sampling inside an interval: the slope of $\nu(\theta)$ rises when $x_0$ becomes smaller.
\begin{center}
\begin{figure}
\includegraphics[width=95mm]{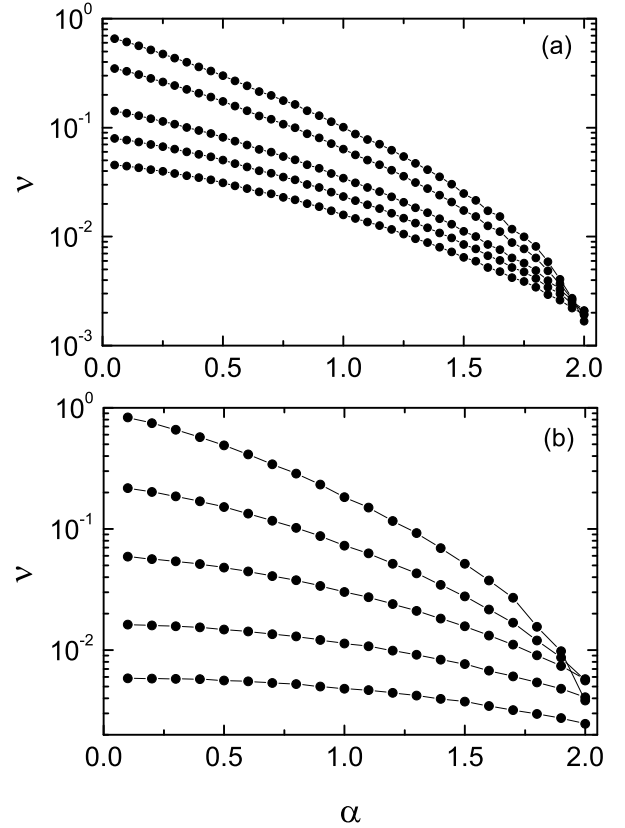}
\caption{Rate $\nu$ as a function of $\alpha$ for $\beta=0.5$ and $L=10$. (a) and (b) correspond to 1D and 2D (isotropic) cases, 
respectively. The curves correspond to the following values of $\theta$ (from top to bottom): 0, -0.2, -0.5, -0.7 and -0.9 (a) 
and 0, -0.5, -1, -1.5 and -1.9 (b).}
\end{figure}
\end{center} 
\begin{center}
\begin{figure}
\includegraphics[width=95mm]{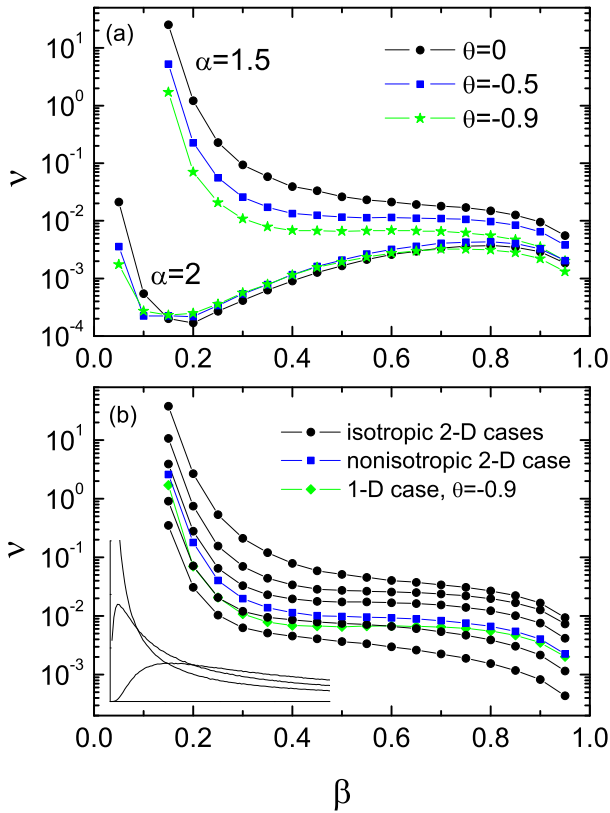}
\caption{(Colour online) $\nu$ as a function of $\beta$ for $\beta=0.5$; the panel (a) corresponds to the 1D problem. The panel (b) 
contains curves calculated for $\alpha=1.5$ and, for the isotropic case, the following values of $\theta$: 
0, -0.5, -0.9, -1.5 and -1.9 (from top to bottom). Inset: Shape of the one-sided L\'evy distribution for $\beta=0.2$, 0.3 and 0.4 
(from up to down at the left-hand side).}
\end{figure}
\end{center} 

The dependence of $\nu$ on $\alpha$ is displayed in Fig.3 for some values of $\theta$. The curves rapidly fall with $\alpha$ -- 
faster than exponentially -- which is the obvious consequence of the rising slope of the PDF asymptotics. 
However, this conclusion has to be qualified if $\theta$ is very negative: the fall of $\nu(\alpha)$ is strongest for $\theta=0$ while 
$\nu(\alpha)$ is flat for $\theta<-1$. The latter observation becomes clear when we consider the Langevin equations corresponding to CTRW, 
Eq.(\ref{lanm}). Then the multiplicative factor reduces the effective jump length when $|{\bf r}|$ is small and $\theta$ negative. 

Fig.4 shows $\nu$ as a function of $\beta$ for both 1D and 2D problems. The dependence $\nu(\beta)$ discloses 
a kind of a plateau for $\beta$ around 0.6 and strongly rises when $\beta$ approaches small values. The latter effect may be 
attributed to the shape of the one-sided L\'evy distribution, Eq.(\ref{lev1s}), which is depicted in the figure and 
exhibits a high maximum near zero when $\beta$ is small. 
If we introduce a cut-off of $w(\xi)$ at small $\xi$ for $\beta<0.4$, which could be reasonable when a system is characterised by 
a low limit of the waiting time, the growth of $\nu(\beta)$ for small $\beta$ does not emerge. 
The above observations are valid for all values of $\theta$ and both dimensionalities; $\nu$ diminishes when $\theta$ becomes more 
negative, in agreement with Fig.2. $\nu$ is largest for the 2D isotropic case and smallest for the nonisotropic case, 
the curve for 1D case lies in-between. For the Gaussian case, 
a deep minimum of $\nu(\beta)$ is observed and the dependence on $\theta$ is weak, except a region of very small $\beta$. 

The above results for the 2D cases have been obtained on the assumption that the jump lengths in both directions are independent. 
If one assumes, instead, that $\boldsymbol\eta(d\tau)$ has a uniform distribution on a circle, $\nu$ becomes systematically 
smaller than that for the symmetric case but the results (not presented) remain qualitatively the same. 

\subsection{Langevin equations and first passage time statistics} 

The system of equations (\ref{lanm}) corresponds, in 1D case, 
to a fractional Fokker-Planck equation with a variable diffusion coefficient \cite{sro15a,kaz}, 
\begin{equation}
\label{glef}
\frac{\partial p(x,t)}{\partial t}={_0}D_t^{1-\beta}\frac{\partial^\alpha}{\partial|x|^\alpha}[|x|^{-\theta\beta} p(x,t)],
\end{equation} 
where the Riemann-Liouville fractional derivative is defined by the expression 
\begin{equation}
\label{rlo}
_0D_t^{1-\beta}f(t)=\frac{1}{\Gamma(\beta)}\frac{d}{dt}\int_0^t dt'\frac{f(t')}{(t-t')^{1-\beta}}  
\end{equation} 
and the Riesz-Weyl derivative by its Fourier transform, 
\begin{equation}
\label{rwder}
\frac{\partial^\alpha}{\partial|x|^\alpha}f(x)={\cal F}^{-1}(-|k|^\alpha{\widetilde f(k)}). 
\end{equation}
Eq.(\ref{glef}), supplemented by the initial and boundary conditions, determines the escape problem. 
On the other hand, we can express the density given by the first equation (\ref{lanm}) as a solution 
of a Fokker-Planck equation which is local in time, 
\begin{equation}
\label{frace}
\frac{\partial p_0(x,\tau)}{\partial \tau}=
\frac{\partial^\alpha[|x|^{-\theta\beta}p_0(x,\tau)]}{\partial|x|^\alpha}, 
\end{equation}  
and integrate over the operational time, 
\begin{equation}
\label{inte}
p(x,t)=\int_0^\infty p_0(x,\tau)h(t,\tau)d\tau. 
\end{equation} 
In the above integral, $h(t,\tau)$ is an inverse subordinator transforming the operational time to the random, physical time. 
More precisely, the random time $t$ is given by an one-sided, 
maximally asymmetric stable L\'evy distribution $h'(\tau,t)=L_\beta(\tau)$, where $0<\beta<1$, which vanishes for $\tau<0$. 
Then $h(t,\tau)=\frac{t}{\beta\tau}L_\beta(\frac{t}{\tau^{1/\beta}})$ \cite{subor}. 

We express the solution of Eq.(\ref{frace}), which satisfies 
the conditions $p(x,0)=\delta(x-x_0)$ ($0<x_0<L$) and $p(0,t)=p(L,t)=0$, in the form, 
\begin{equation}
\label{separ}
p_0(x,\tau)=\sum_n\phi_n(\tau)\psi_n(x). 
\end{equation}
As a first step, let us consider the case $\alpha=2$. Inserting Eq.(\ref{separ}) into Eq.(\ref{frace}) produces the equation \cite{kam1} 
\begin{equation}
  \label{4}
  \frac{\partial^2 x^{-\theta\beta}\psi_n(x)}{\partial x^2}+\lambda_n^2 \psi_n(x)=0,
 \end{equation} 
where $\lambda_n=$const, the solution of which can be expressed in terms of the Bessel function, 
\begin{equation}
  \label{5}
 \psi_n(x)\propto x^{\theta\beta+1/2}J_{1/(\theta\beta+2)}\left(\frac{\sqrt{2}\lambda_n}
  {\theta\beta+2}x^{(\theta\beta+2)/2}\right). 
\end{equation} 
It follows from the boundary conditions that the parameters $\lambda_n$ are given by zeros $\gamma_n$ of the Bessel function, 
$\lambda_n=(\theta\beta+2)\gamma_n/\sqrt{2}L^{(\theta\beta+2)/2}$. Moreover, Eq.(\ref{frace}) yields 
$\phi_n(\tau)\propto \hbox{e}^{-\lambda_n^2\tau}$ and the normalised PDF reads \cite{kam1}, 
\begin{equation}
  \label{8}
\begin{split}
  p_0(x,\tau)&=x^{\theta\beta+1/2}\sum_n C_n J_{1/(\theta\beta+2)}\left(\frac{\gamma_n}
  {L^{(\theta\beta+2)/2}}x^{(\theta\beta+2)/2}\right)\\
  &\times\exp\left(-\frac{(\theta\beta+2)^2\gamma_n^2}{2L^{\theta\beta+2}}\tau\right),
\end{split}
  \end{equation} 
where 
\begin{equation}
  \label{11}
  C_n=\frac{\theta\beta+2}{L^{\theta\beta+2}}\frac{\sqrt{x_0}J_{1/(\theta\beta+2)}\left(\frac{\gamma_n}
  {L^{(\theta\beta+2)/2}}x_0^{(\theta\beta+2)/2}\right)}{[J'_{1/(\theta\beta+2)}(\gamma_n)]^2}. 
\end{equation} 
After performing the integral over $x$, the survival probability in the operational time takes a form 
\begin{equation}
  \label{st}
\begin{split}  
S(\tau)&=\frac{2x_0}{\sqrt{L}}\sum_n\frac{J_{1/(\theta\beta+2)}\left(\frac{\gamma_n}
  {L^{(\theta\beta+2)/2}}x_0^{(\theta\beta+2)/2}\right)}{
\gamma_n J_{-(\theta\beta+1)/(\theta\beta+2)}(\gamma_n)}\\
&\times\exp\left(-\frac{(\theta\beta+2)^2\gamma_n^2}{2L^{\theta\beta+2}}\tau\right)\equiv\sum_nA_n\hbox{e}^{-B_n\tau}.
\end{split}
\end{equation} 
Next we pass to the physical time $t$ and, using Eq.(\ref{inte}), can express the survival probability as a function of this time 
by its Laplace transform, $t\rightarrow u$. Taking into account that 
${\widetilde h}(u,\tau)=u^{\beta-1}\exp(-\tau u^{-\beta})$ we get 
\begin{equation}
\label{sodu}
{\widetilde S}(u)=u^{\beta-1}\sum_n\int_0^\infty A_n\hbox{e}^{-B_n\tau-\tau u^\beta}d\tau=\sum_nA_n\frac{u^{\beta-1}}{u^\beta+B_n} 
\end{equation}
and the inversion of this transform yields the expansion that contains Mittag-Leffler modes, 
\begin{equation}
\label{sodt}
S(t)=\sum_n A_n \hbox{E}_\beta(-B_nt^\beta). 
\end{equation}
The first passage time distribution follows from Eq.(\ref{pfp}) and, inserting a leading term from the asymptotic expansion 
of the Mittag-Leffler function, gives us this distribution for large $t$: 
\begin{equation}
\label{pfpf}
p_{FP}(t)\sim\frac{\beta}{\Gamma(1-\beta)}\sum_n \frac{A_n}{B_n} t^{-1-\beta}. 
\end{equation}

Having the survival probability calculated, we can estimate the escape time by means of the fractional moments 
using Eq.(\ref{mom}). Integration by parts yields 
\begin{equation}
\begin{split}
\label{mom1}
M_\delta&=-\int_0^\infty t^\delta\frac{\partial}{\partial t}\sum_nA_n \hbox{E}_\beta(-B_nt^\beta)dt\\
&=\frac{\delta}{\beta}\int_0^\infty\sum_n A_nB_n^{-\delta/\beta}\xi^{\delta/\beta-1}\hbox{E}_\beta(-\xi)d\xi
\end{split}
\end{equation}
and next we integrate term by term. The integration by parts is valid and the integrated series convergent if $\delta$ satisfies 
conditions $0<\delta<\beta$. Evaluating a Mellin transform from the Mittag-Leffler function yields the final expression, 
\begin{equation}
\label{momf}
M_\delta=\frac{\pi\delta}{\beta}\frac{1}{\Gamma(1-\delta)\sin(\pi\delta/\beta)}\sum_n A_nB_n^{-\delta/\beta}. 
\end{equation}
The sum in Eq.(\ref{momf}) can be exactly evaluated for $\theta=0$ and we put, for simplicity, $x_0=L/2$; then \cite{yus} 
\begin{equation}
\label{anbn}
A_n=\frac{4}{\pi}\frac{(-1)^n}{2n+1}~~~~\hbox{and}~~~~B_n=(2n+1)^2\pi^2/L^2. 
\end{equation}
The moment reads 
\begin{equation}
\label{mom0}
M_\delta=\frac{4\delta}{\beta}\left(\frac{\pi}{L}\right)^{1-a}\frac{2^{-2a}}{\Gamma(1-\delta)\sin(\pi\delta/\beta)}
\left[\zeta(a,1/4)-\zeta(a,3/4)\right], 
\end{equation}
where $a=1+2\delta/\beta$ and $\zeta(s,q)$ stands for a generalised zeta function. 
\begin{center}
\begin{figure}
\includegraphics[width=95mm]{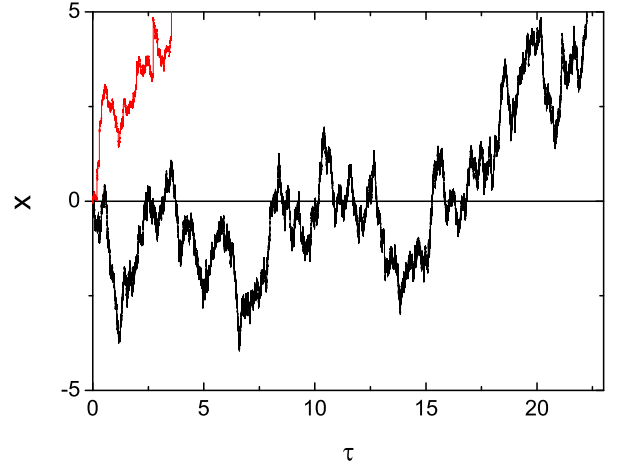}
\caption{(Colour online) Two exemplary trajectories $x(\tau)$ calculated from the Langevin equation for $\beta=0.5$, $L=5$ and two values of $\alpha$: 
2 (black) and 1.5 (the shorter trajectory, marked by the red line).}
\end{figure}
\end{center} 
\begin{center}
\begin{figure}
\includegraphics[width=95mm]{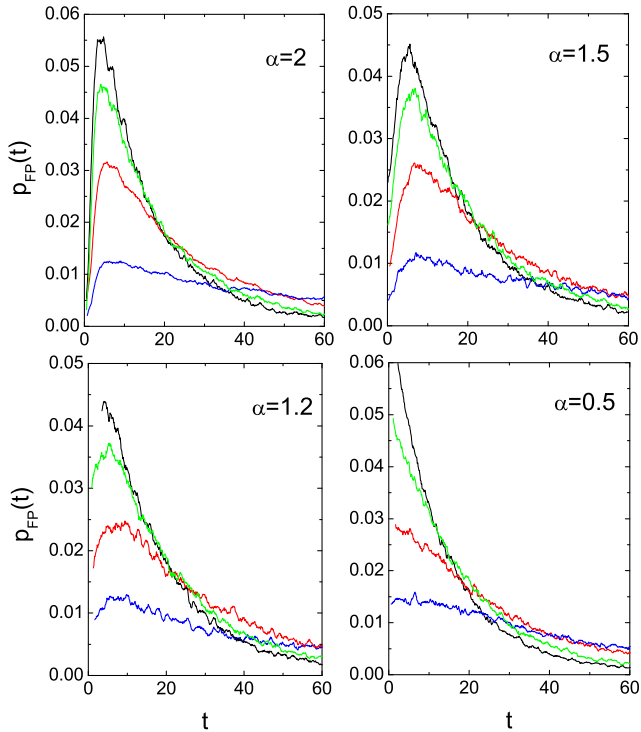}
\caption{(Colour online) First passage time density distributions for $\beta=0.9$, $L=2$, $x_0=0.1$ and some values of $\alpha$. 
The curves in each part of the figure correspond to the following values of $\theta$: 0, -0.2, -0.5 and -0.9 (from top to bottom). 
Each curve was calculated by averaging over $10^5$ trajectories.}
\end{figure}
\end{center} 
\begin{center}
\begin{figure}
\includegraphics[width=95mm]{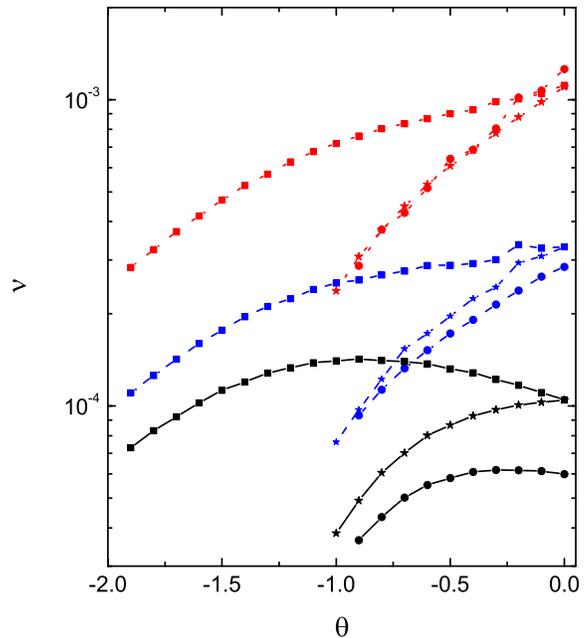}
\caption{(Colour online) $\nu$ as a function of $\theta$ for $\beta=0.9$, $L=2$ and the following values of 
$\alpha$: 2, 1.8 and 1.5 (bunches of curves from bottom to top: solid black, dashed blue and dotted red, respectively). The 1D case 
is marked by points while the 2D case by squares (isotropic case) and stars (nonisotropic case).}
\end{figure}
\end{center} 

According to Eq.(\ref{separ}) and (\ref{inte}), the temporal properties of PDF can be separated 
from its spatial characteristics and the Mittag-Leffler pattern emerges for all $\theta$ and $\alpha\le2$ producing the asymptotics 
$p_{FP}(t)\propto t^{-1-\beta}$. The detailed shape of PDF and its moments can be obtained 
by numerical solutions of Eq.(\ref{lan}) for which we shall apply the same initial and boundary conditions as in Sec.IIIA. 
We discretise both equations and the first equation reads 
\begin{equation}
\label{num1}
{\bf r}(\tau_{n+1})={\bf r}(\tau_n)+\eta_n\Delta\tau^{1/\alpha}, 
\end{equation} 
where $\Delta\tau$ is a time step assumed in all the calculations as $\Delta\tau=10^{-4}$.
The discretisation of the second equation yields 
\begin{equation}
\label{num2}
t(\tau_{n+1})=t(\tau_n)+g[x(\tau_n)]\xi_n\Delta\tau^{1/\beta} 
\end{equation} 
and the random numbers $\eta_n$ and $\xi_n$ are sampled from the stable distributions $L_\alpha$ and $L_\beta$, respectively. 
Eq.(\ref{num1}) determines the evolution in the operational time and evolution in the random time $t$ follows from the relation 
between those times, Eq.(\ref{num3}).  
Typical examples of trajectories illustrating the evolution in the operational time are presented in Fig.5; 
the evolution terminates when the barrier at $|x|=L$ is reached. The trajectories correspond 
to both the Gaussian case and the L\'evy flights; they look differently: in the latter case a monotonic growth of the distance prevails 
while the Gaussian case exhibits an oscillatory structure with many returns to the origin. Obviously, since the evolution proceeds 
in the operational time, the above observations are independent of a specific trap structure. In many dimensions, 
the shape of the trajectory does not depend on $g(x)$ and the trap structure influences only the time characteristics. 
The fact that the same sites are visited many times has important consequences if, in the disordered systems, 
correlations of the random walk with the disorder are taken into account (quenched disorder): the renewal theory (valid 
for the annealed disorder) does not apply and the subordination technique 
breaks down for the one-dimensional, Gaussian case \cite{bar}; as a consequence, the transport properties for both disorders are 
different \cite{bou}. Fig.5 suggests that those properties could be more similar for the L\'evy flights. 

The trajectory simulations yield the first passage time and determine the density distributions. 
They are presented in Fig.6 for 1D case and some values of $\alpha$ and $\theta$. 
The density $p_{FP}(t)$ appears very sensitive to $\theta$; since the negative values of $\theta$ 
result in longer trapping time, $p_{FP}(t)$ is suppressed at small $t$ while the tails are raised up, 
compared to the homogeneous case. The shapes of the presented $p_{FP}(t)$, corresponding to the same $\theta$, are similar 
except the case $\alpha=0.5$ when the region of very short escape times is pronounced because of large values of the random force. 

The averaging over the numerically evaluated $p_{FP}(t)$ produces the fractional moments and the rate $\nu$. 
The dependence of $\nu$ on the system parameters appears similar to that presented in Sec.IIIA for 
the random walk simulations and we summarise the most important results of that 
analysis in Fig.7 which corresponds to Fig.2. $\nu$ rises with $\theta$ for all the cases except $\alpha=2$ and the growth 
is weakest for the isotropic 2D case. Moreover, the results for 2D nonisotropic and the 1D cases are similar 
unless $\alpha=2$. 

Presence of the Mittag-Leffler function in problems related to escape phenomena is also known for deterministic systems: 
it emerges in a description of the first passage time for intermittent weakly chaotic systems with invariant
density similar to Eq.(\ref{godr}) \cite{vene}.

\section{Summary and conclusions}

We have considered transport properties in a stochastic system characterised by long resting times that are random but biased by 
the medium structure: the medium contains traps, responsible for the rests and memory effects. Traps are nonhomogeneously distributed 
which has been taken into account by introducing -- in the CTRW description -- a position-dependent trap density.
The assumed power-law form of this density has been interpreted as a result of the underlying self-similar structure. 
The jump lengths are governed by the general symmetric stable distribution. 
On the other hand, a stochastic dynamics, as a continuous counterpart of CTRW, has been formulated in terms 
of the subordination technique where the random time generator is position-dependent and the resulting system of the Langevin 
equations resolves itself to a subordination of a multiplicative process to the random time. This approach allows us to take into account 
the nonhomogneity effects in a simple way: by a fractional Fokker-Planck equation. 

The heterogeneity of the medium structure influences time characteristics of the escape from a given area, in particular 
the first passage time PDF and its moments. To quantify the escape speed we have evaluated the escape rate $\nu$, namely 
the moment of the order -1, which is proportional to the effective particle velocity in the bulk, as a function of the stability index $\alpha$, 
memory parameter $\beta$ and the trap structure parameter $\theta$. $\nu$ decreases with $\alpha$, $\beta$ and rising nonhomogeneity 
(more negative $\theta$), but only for $\alpha<2$. The Gaussian case is special since the dependences $\nu(\theta)$ and $\nu(\beta)$ are 
not monotonous which effect can be attributed to multiple returns to the origin. Analysis of 2D case shows that the escape speed 
is sensitive to the isotropy of the trap structure: $\nu$ differs for the nonisotropic case from that for the isotropic one though 
the underlying fractal structure has the same capacity dimension for a given $\theta$. Another consequence of the variable trap density 
is the sensitivity of the dynamics to the initial position which reflects the fact that the trapping is strongest near the origin. 
The variable and diminishing density $g(x)$ substantially lowers the central part of the first passage time PDF making the escape time larger. 
On the other hand, the strong memory (small $\beta$) may invoke a rapid escape which is a consequence of the enhancement of 
the waiting time PDF near the origin and occurs for any $\alpha$ and $\theta$.


\begin{thebibliography}{99}

\bibitem{met}
R. Metzler and J. Klafter, Phys. Rep. {\bf 339}, 1 (2000). 

\bibitem{zab}
V. Zaburdaev, S. Denisow, and J. Klafter, Rev. Mod. Phys. {\bf 87}, 483, (2015). 

\bibitem{kam}
A. Kami\'nska and T. Srokowski, Phys. Rev. E  {\bf 69}, 062103 (2004). 

\bibitem{tak}
{\it Cooperative Dynamics in Complex Systems}, edited by H. Takayama (Springer, Berlin, 1989). 

\bibitem{bou}
J.-P. Bouchaud and A. Georges, Phys. Rep. {\bf 195}, 127 (1990). 

\bibitem{zas}
G. M. Zaslavski, Phys. Rep. {\bf 371}, 461 (2002). 

\bibitem{bro1}
D. Brockmann, L. Hufnagel, and T. Geisel, Nature {\bf 439}, 462 (2006). 

\bibitem{sro06}
T. Srokowski and A. Kami\'nska, Phys. Rev. E {\bf 74}, 021103 (2006). 

\bibitem{shles}
{\it L\'evy flights and related topics in physics}, 
edited by M. F. Shlesinger, G. M. Zaslavsky, and J. Frisch (Springer Verlag, Berlin, 1995).

\bibitem{barn}
{\it L\'evy processes: Theory and applications}, 
edited by O. E. Barndorff-Nielsen, T. Mikosch, and S. I. Resnick
(Birkh\"auser, Boston, 2001). 

\bibitem{tall}
K. T. Tallakstad, R. Toussaint, S. Santucci, and K. J. M\aa l\o y, Phys. Rev. Lett. {\bf 110}, 145501 (2013). 

\bibitem{bro}
D. Brockmann and T. Geisel, Phys. Rev. Lett. {\bf 90}, 170601 (2003). 

\bibitem{sti}
A. V. Chechkin, R. Gorenflo, and I. M. Sokolov, J. Phys. A {\bf 38}, L679 (2005); 
B. A. Stickler and E. Schachinger, Phys. Rev. E {\bf 83}, 011122 (2011); 
{\it ibid} {\bf 84}, 021116 (2011). 
 
\bibitem{subor}
H. C. Fogedby, Phys. Rev. E {\bf 50}, 1657 (1994); E. Barkai, Phys. Rev. E {\bf 63}, 046118 (2001); 
A. Piryatinska, A. I. Saichev, and W. A. Woyczynski, Physica A {\bf 349}, 375 (2005). 

\bibitem{sro14} 
T. Srokowski, Phys. Rev. E {\bf 89}, 030102(R) (2014). 

\bibitem{sro15}
T. Srokowski, Phys. Rev. E {\bf 91}, 052141 (2015). 

\bibitem{han}
P. H\"anggi, P. Talkner, and M. Borkovec, Rev. Mod. Phys. {\bf 62}, 251 (1990). 

\bibitem{kore}
T. Koren, M. A. Lomholt, A. V. Chechkin, J. Klafter, and R. Metzler, 
Phys. Rev. Lett. {\bf 99}, 160602 (2007). 

\bibitem{dyb}
B. Dybiec, E. Gudowska-Nowak, and P. H\"anggi, Phys. Rev. E {\bf 73}, 046104 (2006). 

\bibitem{dyb1}
B. Dybiec, E. Gudowska-Nowak, and P. H\"anggi, Phys. Rev. E {\bf 75}, 021109 (2007). 

\bibitem{szcz}
K. Szczepaniec and B. Dybiec, Phys. Rev. E {\bf 89}, 042138 (2014). 

\bibitem{luo}
L. Luo and L.-H. Tang, Phys. Rev. E {\bf 92}, 042137 (2015). 

\bibitem{yus} 
S. B. Yuste and K. Lindenberg, Phys. Rev. E {\bf 69}, 033101 (2004). 

\bibitem{uwa}
For some problems involving L\'evy flights, the jumping time may be small, compared to the waiting time, and can be neglected; 
that is the case, e.g., for the population movement \cite{bro1}. 

\bibitem{man}
R. N. Mantegna and H. E. Stanley, Phys. Rev. Lett. {\bf 73}, 2946 (1994); I. Koponen, Phys. Rev. E {\bf 52}, 1197 (1995). 

\bibitem{schn}
W. R. Schneider, in {\it Stochastic Processes in Classical and Quantum Systems, 
Lecture Notes in Physics}, edited by S. Albeverio, G. Casati, D. Merlini 
(Springer, Berlin, 1986), Vol. 262. 

\bibitem{lich}
A. J. Lichtenberg and M. A. Lieberman, {\it Regular and Chaotic Dynamics}
(Springer -- Verlag, New York, 1992). 

\bibitem{hav}
S. Havlin and D. Ben-Avraham, Adv. Phys. {\bf 51}, 187 (2002). 

\bibitem{met1} 
The problem of a random walk of a particle hopping among fractally distributed sites and experiencing long rests 
resolves itself to a Fokker-Planck equation with a fractional time derivative and a variable diffusion coefficient, cf., 
R. Metzler, W. G. Gl\"ockle, and T. F. Nonnenmacher, Physica A {\bf 211}, 13 (1994). 

\bibitem{wer}
A. Janicki and A. Weron, {\it Simulation and Chaotic Behavior of
$\alpha$-Stable Stochastic Processes} (Marcel Dekker, New York, 1994). 

\bibitem{kaz}
R. Kazakevi\^cius and J. Ruseckas, Physica A {\bf 438}, 210 (2015). 

\bibitem{sro15a}
T. Srokowski, Phys. Rev. E {\bf 92}, 012125 (2015). 

\bibitem{red}
S. Redner, {\it A guide to first-passage processes}, Cambridge University Press, 2001. 

\bibitem{paly}
V. V. Palyulin, A. V. Chechkin, and R. Metzler, Proc. Natl. Acad. Sci. USA {\bf 111}, 2931 (2014); 
V. V. Palyulin, A. V. Chechkin, and R. Metzler, J. Stat. Mech. (2014) P11031. 

\bibitem{zum} 
G. Zumofen and J. Klafter, Phys. Rev. E {\bf 51}, 2805 (1995). 

\bibitem{kam1}
A. Kami\'nska and T. Srokowski, Acta Phys. Pol. {\bf 38}, 3119 (2007). 

\bibitem{bar}
S. Burov and E. Barkai, Phys. Rev. E {\bf 86}, 041137 (2012). 

\bibitem{vene} 
P. Naze and R. Venegeroles, Phys. Rev. E {\bf 90}, 042917 (2014); R. Venegeroles, Phys. Rev. E {\bf 91}, 062914 (2015); 
R. Venegeroles, Phys. Rev. E {\bf 86}, 021114 (2012). 
 
\end{thebibliography}
\end{document}